\documentclass[a4paper,11pt]{article}
\usepackage{pos}
\usepackage{amsmath}
\usepackage{placeins}
\usepackage{multirow}
\usepackage{tabularx}
\usepackage{hyperref} 

\title{Antiheavy-Antiheavy-Light-Light Four-Quark Bound States}

\author*[a]{Martin Pflaumer}
\author[b,c]{Constantia Alexandrou}
\author[c]{Jacob Finkenrath}
\author[d]{Theodoros Leontiou}
\author[e]{Stefan Meinel}
\author[a,f]{Marc Wagner}

\affiliation[a]{Institut f\"ur Theoretische Physik, Goethe-Universit\"at Frankfurt am Main, \\ Max-von-Laue-Stra{\ss}e 1, D-60438 Frankfurt am Main, Germany}

\affiliation[b]{Department of Physics, University of Cyprus,\\
	20537 Nicosia, Cyprus}

\affiliation[c]{Computation-based Science and Technology Research Center, The Cyprus Institute,\\
	20 Konstantinou Kavafi Street, 2121 Nicosia, Cyprus}

\affiliation[d]{Department of Mechanical Engineering, Frederick University,\\
	1036 Nicosia, Cyprus}

\affiliation[e]{Department of Physics, University of Arizona,\\
	Tucson, AZ 85721, USA} 

\affiliation[f]{Helmholtz Research Academy Hesse for FAIR,\\
	Campus Riedberg, Max-von-Laue-Stra{\ss}e 12, D-60438 Frankfurt am Main, Germany}

\emailAdd{pflaumer@itp.uni-frankfurt.de}
\emailAdd{alexand@ucy.ac.cy}
\emailAdd{j.finkenrath@cyi.ac.cy}
\emailAdd{t.leontiou@frederick.ac.cy}
\emailAdd{smeinel@arizona.edu}
\emailAdd{mwagner@itp.uni-frankfurt.de}

\abstract{We present our recent results on antiheavy-antiheavy-light-light tetraquark systems using lattice QCD. Our study of the $ \bar{b}\bar{b}us $ four-quark system with quantum numbers $ J^P=1^+ $ and the $ \bar{b}\bar{c}ud $ four-quark systems with $ I(J^P)=0(0^+) $ and $ I(J^P)=0(1^+) $ utilizes scattering operators at the sink to improve the extraction of the low-lying energy levels. We found a bound state for $ \bar{b}\bar{b}us $ with $ E_{\textrm{bind},\bar{b}\bar{b}us} = (-86 \pm 22 \pm 10)\,\textrm{MeV} $, but no indication for a bound state in both $ \bar{b}\bar{c}ud $ channels. Moreover, we show preliminary results for $ \bar{b}\bar{b}ud $  with $ I(J^P)=0(1^+) $, where we used scattering operators both at the sink and the source. We found a bound state and determined its infinite-volume binding energy with a scattering analysis, resulting in $ E_{\textrm{bind},\bar{b}\bar{b}ud} =(-103 \pm 8 )\,\textrm{MeV} $.}

\FullConference{%
The 39th International Symposium on Lattice Field Theory,\\
8th-13th August, 2022,\\
Rheinische Friedrich-Wilhelms-Universität Bonn, Bonn, Germany
}

\newcolumntype{C}[1]{>{\centering\arraybackslash}m{#1}}

\begin{document}
\maketitle

\section{Introduction}

Within the last years, a large number of four-quark states have been discovered experimentally as well as predicted theoretically. There are also a lot of ongoing activities and more discoveries are expected in the near future. An example of a recent discovery is the $ T_{cc}^+$  $(cc\bar{u}\bar{b})$ tetraquark state which has been found as a weakly bound tetraquark state by the LHCb collaboration \cite{LHCb:2021auc,LHCb:2021vvq}. A subsequent lattice study investigating the same $cc\bar{u}\bar{b}$ state predicts a virtual bound state slightly below the $ DD^* $ threshold \cite{Padmanath:2022cvl}.
In this work, we also use lattice QCD and study similar antiheavy-antiheavy-light-light four-quark states $ \bar{Q}\bar{Q}'qq' $, where at least one of the heavy quarks is a bottom quark, i.e. $ \bar{Q}=\bar{b} $.

Such systems were first investigated with lattice methods using the Born-Oppen\-heimer approximation. Those studies predicted a hadronically stable $ \bar{b}\bar{b}ud $ tetraquark with binding energy $ \approx60 \ldots 90 \,\textrm{MeV}$ below the $ BB^* $ threshold in the $ I(J^P)=0(1^+) $ channel \cite{Brown:2012tm,Bicudo:2012qt,Bicudo:2015kna,Bicudo:2015vta,Bicudo:2016ooe,Bicudo:2021qxj}. Additionally, a $ \bar{b}\bar{b}ud $ resonance with quantum numbers $ I(J^P)=0(1^-) $ was found $ \approx 20\,\textrm{MeV} $ above the $ BB $ threshold with decay width $ \Gamma \approx 100\,\textrm{MeV} $ \cite{Bicudo:2017szl}. Furthermore, full lattice QCD studies using NRQCD for the bottom quarks confirm the existence of the hadronically stable $ \bar{b}\bar{b}ud $ tetraquark, predicting a binding energy of magnitude $ 130\ldots 190 \,\textrm{MeV} $ \cite{Francis:2016hui,Junnarkar:2018twb,Leskovec:2019ioa}.
Moreover, extensive studies of $ \bar{b}\bar{b}us $ and $ \bar{b}\bar{c}ud $ four-quark states have been carried out within the same lattice setups. In the $ \bar{b}\bar{b}us $ case, a hadronically stable tetraquark with binding energy of $ \approx 80\,\textrm{MeV} $ has been predicted, while for the $ \bar{b}\bar{c}ud $ systems, the situation is less clear and results are still inconclusive \cite{Francis:2017bjr,Junnarkar:2018twb,Francis:2018jyb,Hudspith:2020tdf,Mohanta:2020eed,Mathur:2021gqn,Meinel:2022lzo}.

In this paper, we summarize our recent and ongoing activities with improved operator bases using scattering operators at the sink for 
$ \bar{b}\bar{b}us $ and $ \bar{b}\bar{c}ud $ as well as using scattering operators both at the sink and the source for
$ \bar{b}\bar{b}ud $.
%
%
\section{Investigation of $ \bar{b}\bar{b}us $ and $ \bar{b}\bar{c}ud $ four-quark states using scattering operators at the sink\label{Sec:bbus_bcud}}

In previous lattice QCD studies focusing on $ \bar{b}\bar{b}us $ and $ \bar{b}\bar{c}ud $ four-quark systems, only local operators have been utilized. With local operators we refer to operators in which all four quarks are centered at the same point in space with total momentum projected to zero. Here we extend the operator basis by scattering operators, which resemble two spatially separated mesons, i.e.\ the meson momenta are separately projected to zero.
If a bound state exists in a specific channel, we expect that local operators generate sizable overlaps to that bound state. Moreover, scattering operators allow to also resolve two-meson scattering states, which are expected to be close to possibly existing bound states.
In recent work \cite{Leskovec:2019ioa,Meinel:2022lzo}, we highlighted the importance of scattering operators by demonstrating that the ground-state energy decreases for all systems significantly, if scattering operators are included.

In this section, we summarize our recent publication \cite{Meinel:2022lzo}, where we studied quantum numbers $ J^P=1^+ $ 
for quark flavors $ \bar{b}\bar{b}us $ and the two channels $ I(J^P)=0(0^+) $ and $ I(J^P)=0(1^+) $ for quark flavors $ \bar{b}\bar{c}ud $. For each system, we consider several local mesonic operators and one local diquark-antidiquark operator both at the sink and the source as well as several scattering operators only at the sink. For more details about the operators, we refer to Ref.~\cite{Meinel:2022lzo}.

\subsection{Lattice setup \label{Sec:RBCSetup}}

We used gauge link configurations generated by the RBC and UKQCD collaboration with $ 2+1 $ flavours of domain-wall fermions and Iwasaki gauge action \cite{Aoki:2010dy,Blum:2014tka}. We used five different ensembles which differ in the lattice spacing, the spatial extent and the pion mass, with one ensemble at the physical pion mass. The parameters of the five ensembles can be found in Table~1 of Ref.~\cite{Meinel:2022lzo}.

We used smeared point-to-all propagators for all quark flavors. The bottom quarks are treated in the framework of NRQCD \cite{Thacker:1990bm,Lepage:1992tx}, whereas we applied a relativistic heavy quark action for the charm quarks \cite{Brown:2014ena}. Due to the use of point-to-all propagators, scattering operators could only be implemented at the sink. Consequently, our correlation matrices are non-square.


\subsection{Existence of a hadronically stable $ \bar{b}\bar{b}us $ four-quark state with $ J^P = 1^+ $}

\begin{figure}[h]
	\centering
	\includegraphics[width=0.49\linewidth]{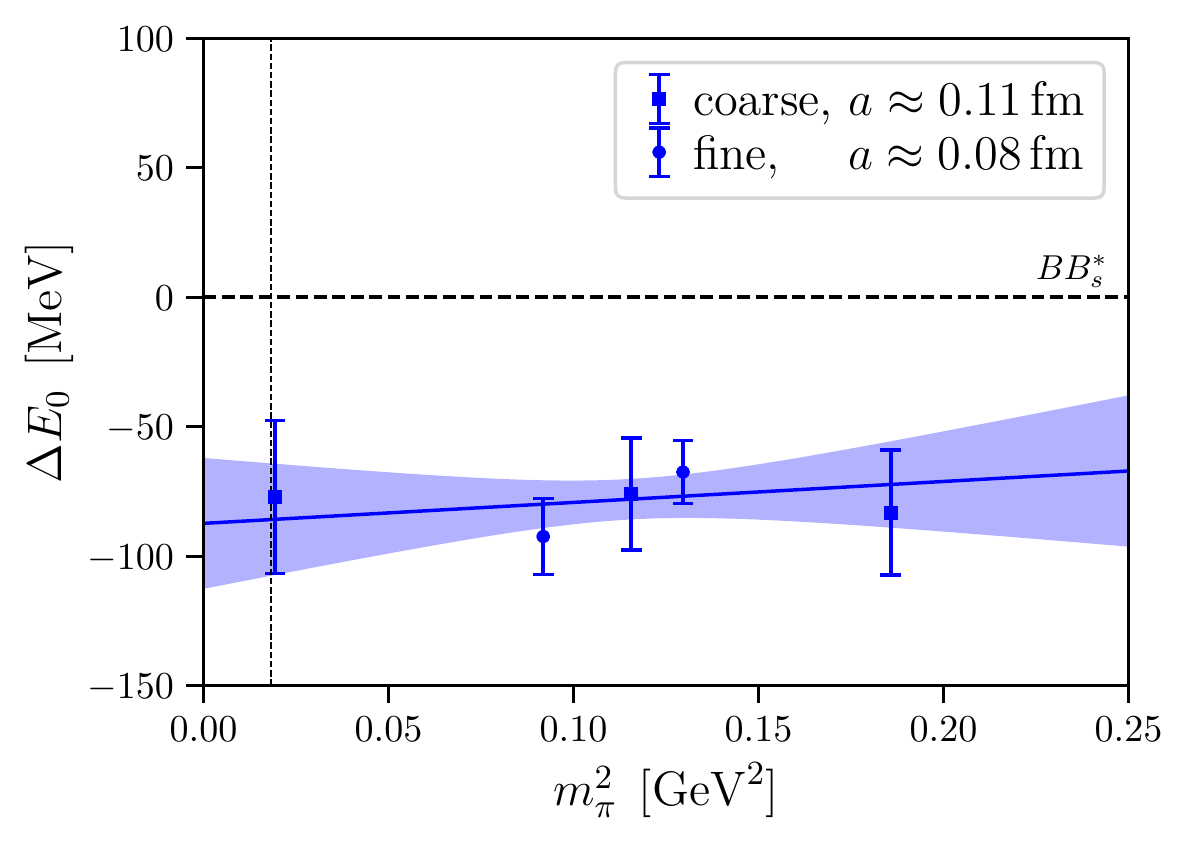}
	\caption{Ground-state energy as function of the squared pion mass for the $ \bar{b}\bar{b}us $ system. The blue line and error band indicates the fit to the data points and linear extrapolation to the physical pion mass at $ m_{\pi,\textrm{phys}} = 135\,\textrm{MeV} $. \label{fig:bbus_res}}
\end{figure}

We determined the ground-state energy for all five ensembles discussed in Sec.~\ref{Sec:RBCSetup}. In Fig.~\ref{fig:bbus_res}, we plot the ground-state energy as function of the squared pion mass relative to the lowest two-meson threshold $ BB_s^* $. For all ensembles this energy is clearly below the threshold, which provides strong evidence for a hadronically stable tetraquark state. Additionally, we performed  a chiral extrapolation to the physical pion mass, which yields a binding energy of 
\begin{equation}
	\Delta E_0 (m_{\pi,\textrm{phys}}) = (-86 \pm 22 \pm 10)\,\textrm{MeV}.
\end{equation}

We also computed the overlaps of the trial states to the low lying energy eigenstates. In particular we found that one of our local operators generates a large ground-state overlap but only little overlap to excited states. This supports our interpretation of the ground-state as a hadronically stable tetraquark. For more details see Ref.~\cite{Meinel:2022lzo}.
%


\subsection{$ \bar{b}\bar{c}ud $ four-quark states with $ I(J^P)=0(0^+) $ and $ I(J^P)=0(1^+) $}
\begin{figure}[h]
	\centering
	\includegraphics[width=0.49\linewidth]{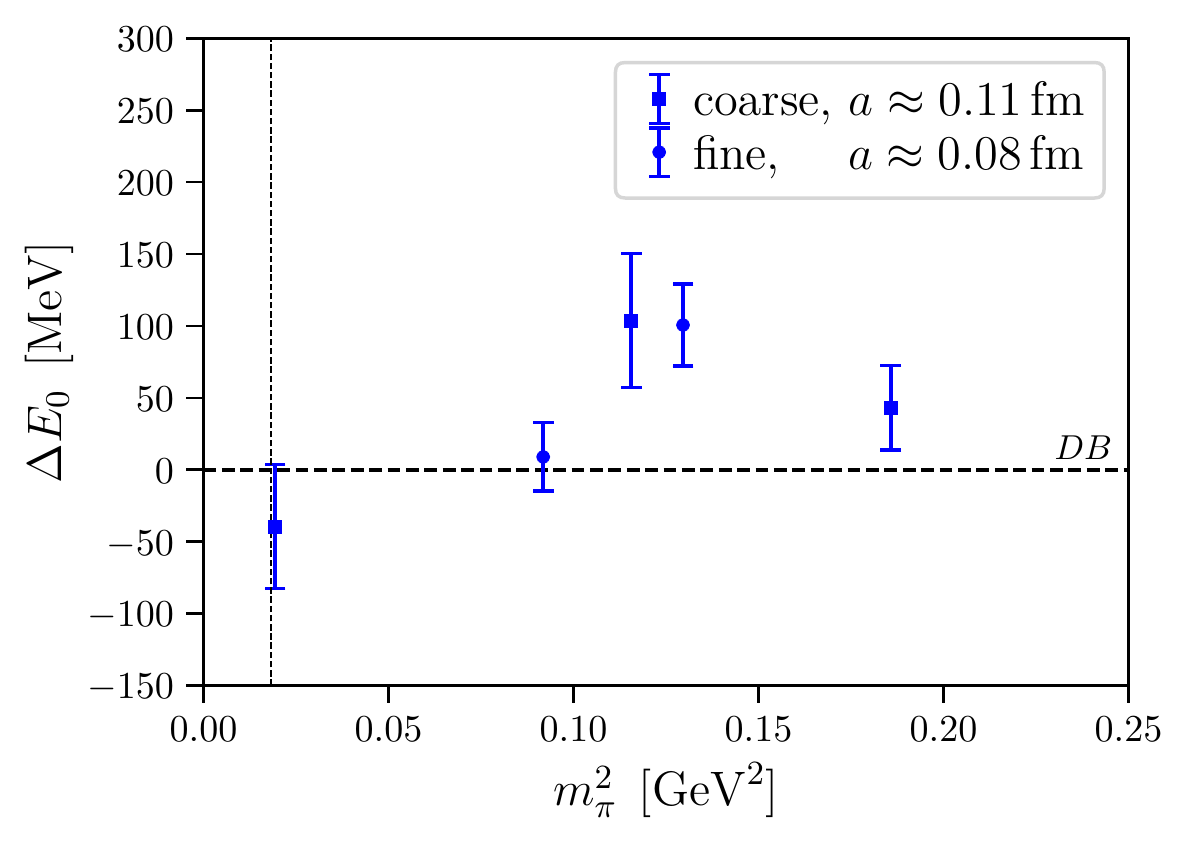}
	\includegraphics[width=0.49\linewidth]{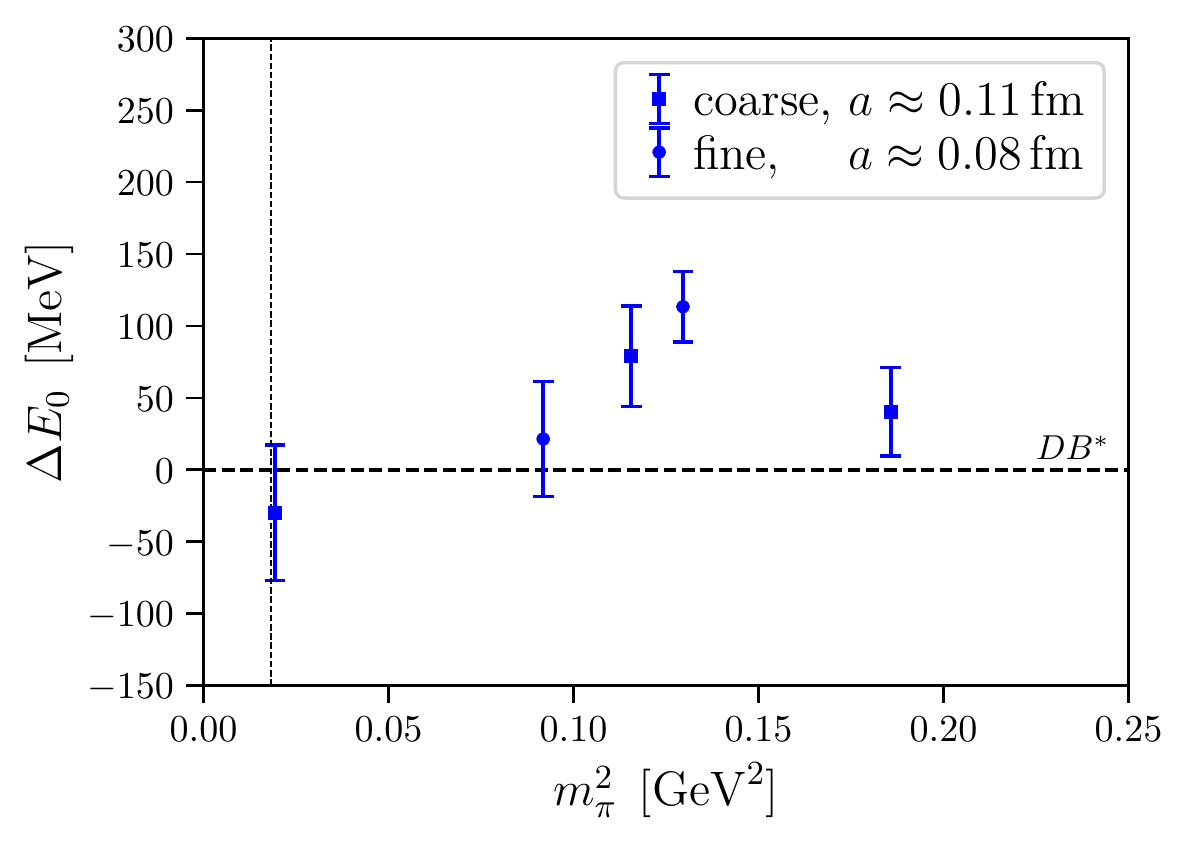}
	\caption{Ground-state energy as function of the squared pion mass for the $ \bar{b}\bar{c}ud $ system with $ I(J^P)=0(0^+) $ (left) and with $ I(J^P)=0(1^+) $ (right). \label{fig:bcud_res}}
\end{figure}

We present the results for the two $ \bar{b}\bar{c}ud $ channels in Fig.~\ref{fig:bcud_res}. In both cases we find the lowest energy level  above or in agreement with the lowest associated two-meson threshold. Accordingly, there is no indication that a stable tetraquark state exists in either of the two channels.

As before, we also computed the overlap factors. The results suggest that the ground-states are two-meson scattering states rather than tetraquarks (see Ref.~\cite{Meinel:2022lzo} for details).
%


\section{Investigation of a $ \bar{b}\bar{b}ud $ four-quark state using scattering operators both at the sink and at the source}

We have studied the $ \bar{b}\bar{b}ud $ system with $ I(J^P)=0(1^+) $ already some time ago using the same setup and techniques as discussed in Sec.~\ref{Sec:bbus_bcud}, i.e.\ employing scattering operators only at the sink \cite{Leskovec:2019ioa}. Now we are in the process of improving these calculations by including the scattering operators also at the source.
The full implementation of scattering operators is highly beneficial for reliably extracting the low-lying energy levels \cite{Wagner:2022bff}. Moreover, a rigorous scattering analysis using L\"uscher's method, enabled by such a complete implementation, is essential for the study of other tetraquark systems closer to or above strong-decay thresholds.

\subsection{Lattice setup}

This calculation is carried out on gauge link configurations generated by the MILC collaboration with $ 2+1+1 $ flavors of HISQ fermions \cite{MILC:2012znn}. For the valence quarks we use a Wilson-clover action following the approach in Refs.~\cite{Bhattacharya:2015wna,Gupta:2018qil}. We list the parameters of the six ensembles used in this computation in Table~\ref{tab:configurations_MILC}. Note that we have three ensembles with almost the same lattice spacings and pion masses but different spatial extents (a12m220S, a12m220, a12m220L) which allow us to perform a high-quality scattering analysis based on finite-volume energy levels from three different spatial volumes.
\begin{table}[h]
	\centering
	\begin{tabular}{cccccc}
		\hline\hline
		Ensemble & $ N_s^3\times N_t $ & $ a $ [fm] & $ m_{\pi}^{\textrm{(sea)}}\,\textrm{[MeV]} $ & $ m_{\pi}^{\textrm{(val)}}\,\textrm{[MeV]} $ & $ N_{\textrm{conf}} $ \\ \hline
		a12m310  & $ 24^3 \times 64 $  & 0.1207(11) &    $ 305.3(4) $     &   $ 310.2(2.8) $    &   $ 1053 $   \\
		a12m220S & $ 24^3 \times 64 $  & 0.1202(12) &    $ 218.1(4) $     &   $ 225.0(2.3) $    &   $ 1020 $   \\
		a12m220  & $ 32^3 \times 64 $  & 0.1184(10) &    $ 216.9(2) $     &   $ 227.9(1.9) $    &   $ 1000 $   \\
		a12m220L & $ 40^3 \times 64 $  & 0.1189(09) &    $ 217.0(2) $     &   $ 227.6(1.7) $    &   $ 1030 $   \\
		a09m310  & $ 32^3 \times 96 $  & 0.0888(08) &    $ 312.7(6) $     &   $ 313.0(2.8) $    &   $ 1166 $   \\
		a09m220  & $ 48^3 \times 96 $  & 0.0872(07) &    $ 220.3(2) $     &   $ 225.9(1.8) $    &   $ 657 $    \\ \hline\hline
	\end{tabular}
	\caption{\label{tab:configurations_MILC}Gauge link ensembles \cite{MILC:2012znn} used in this work. $N_s$, $N_t$: number of lattice sites in spatial and temporal directions; $a$: lattice spacing; $m_{\pi}^{\textrm{(sea)}}$: sea quark pion mass; $m_{\pi}^{\textrm{(val)}}$: valence quark pion mass; $N_{\textrm{conf}}$: number of gauge link configurations.}
\end{table}

We use smeared point-to-all propagators for diagrams with local operators at the source and stochastic timeslice-to-all propagators for diagrams with scattering operators at the source. As before, the bottom quarks are treated in the framework of NRQCD.


\subsection{Existence of a hadronically stable $ \bar{b}\bar{b}ud $ four-quark state with $ I(J^P)=0(1^+) $}

We use the same operator basis as in our previous work \cite{Leskovec:2019ioa}. This time, however, we compute the full squared correlation matrix. The lowest finite-volume energy levels are determined by solving a standard generalized eigenvalue problem on all six available ensembles. To determine the infinite-volume ground-state energies on each of the ensembles a09m220, a09m310 and a12m310 and on the set of three ensembles with similar $ a $ and $ m_\pi $ (a12m220S, a12m220, a12m220L), we perform scattering analyses including the two lowest energy levels from each ensemble. 

\begin{figure}[h]
	\centering
	\raisebox{-0.5\height}{\includegraphics[width=0.4\linewidth]{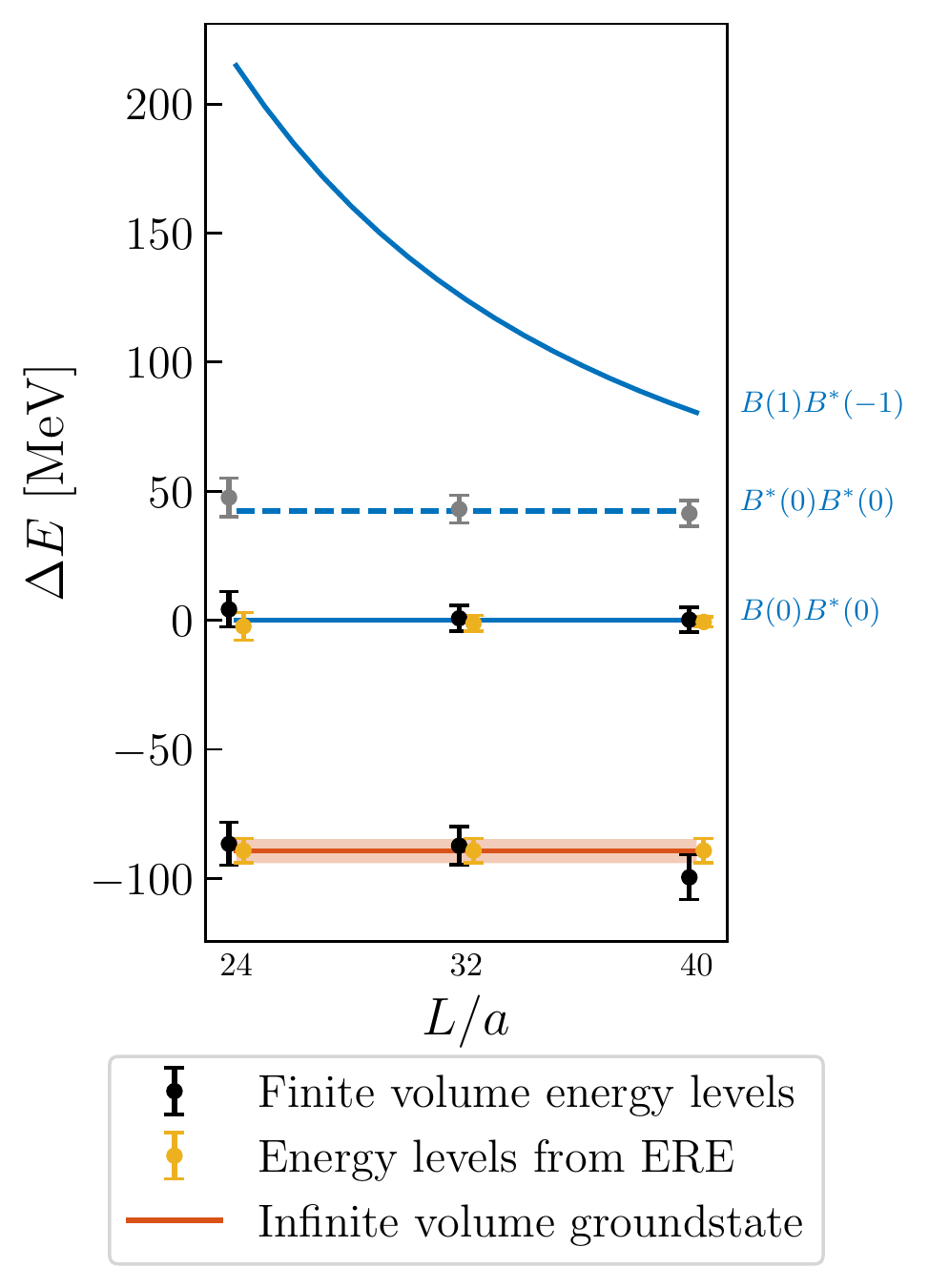}}
	\raisebox{-0.5\height}{\includegraphics[width=0.5\linewidth]{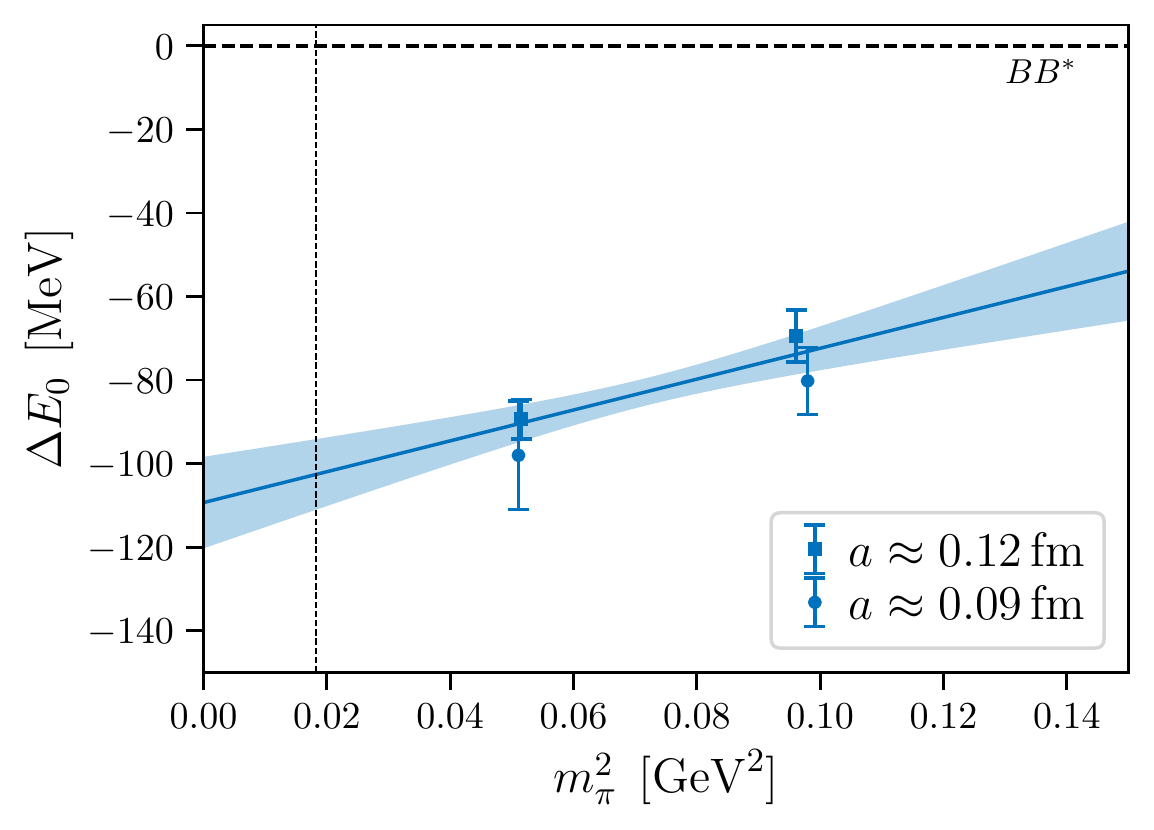}}
	\caption{\label{fig:scattAnalys}$ \bar{b}\bar{b}ud $ system with $ I(J^P)= 0(1^+) $. Left: Finite-volume energy levels for the ensembles a12m220S, a12m220 and a12m220L (black data points are used in the scattering analysis).
	The yellow points are the energy levels obtained by the ERE fit and the red horizontal line represents the resulting infinite-volume ground-state energy. Right: Ground-state energy as function of the squared pion mass. The blue line and error band indicates the fit and linear extrapolation to the physical point at $ m_{\pi,\textrm{phys}} = 135\,\textrm{MeV} $.}
\end{figure}

We define the scattering momenta $ k_n $ ($ n=0,1 $) for each ensemble via
\begin{equation}
	k_n^2 = \frac{\Delta E_n (\Delta E_n + 2m_{B,\textrm{kin}}) (\Delta E_n + 2m_{B^*,\textrm{kin}}) (\Delta E_n + 2m_{B,\textrm{kin}} + 2m_{B^*,\textrm{kin}})}{4(\Delta E_n + m_{B,\textrm{kin}} + m_{B^*,\textrm{kin}})^2}
\end{equation}
with $ \Delta E_n = E(n) - E_B - E_B^* $, where $ E(n) $ is the $ n $-th energy level of the $ \bar{b}\bar{b}ud $ system and $ E_B $ and $ E_B^* $ are the lattice energies of the $ B $- and $ B^* $-meson. The kinetic meson masses $ m_{B,\textrm{kin}} $ and $ m_{B^*,\textrm{kin}} $ are calculated in the same way as described in Sec.~IV of Ref.~\cite{Leskovec:2019ioa}.\\
The finite-volume scattering momenta are related to the infinite-volume $ S $-wave phase shift \cite{Luscher:1990ux} by 
\begin{equation}
	\cot(\delta_0(k_n)) = \frac{2 Z_{00}(1; (k_n L /2\pi)^2)}{\pi^{1/2}k_n L}.
\end{equation}
By parametrizing the scattering phase shift via the effective-range-expansion (ERE),
\begin{equation}
	k \cot \delta_0(k) = \frac{1}{a_0} + \frac{1}{2} r_0 k^2,
	\label{eq:scattCond}
\end{equation}
we relate the parameters $ a_0 $ and $ r_0 $ to finite-volume energy levels $ E_{\textrm{par}}(L,n;a_0, r_0) $, $ n=0,1 $. This allows us to define and to minimize the $ \chi^2 $ function
\begin{equation}
	\chi^2 = \sum_{L} \sum_{n,n'} \Big(E(L,n) - E_{\textrm{par}}(L,n;\{a_i\})\Big) \mathbb{C}^{-1}(L,n,n') \Big(E(L,n') - E_{\textrm{par}}(L,n';\{a_i\})\Big),
\end{equation} 
where $ E(L,n) $ is the $ n $-th lattice energy level for lattice extent $ L $ and $ \mathbb{C}^{-1} $ is the corresponding inverse covariance matrix. Finally, the infinite-volume ground-state energy is obtained by determining the pole of the scattering amplitude
\begin{equation}
	T_0(k) = \frac{1}{\cot\delta_0(k) -i}.
\end{equation}
We show in Fig.~\ref{fig:scattAnalys} (left) the lowest finite-volume energy levels for three different lattice sizes at $ a\simeq 0.12 \,\textrm{fm} $ and $ m_{\pi} \simeq 220\,\textrm{MeV} $ as well as the corresponding ERE fit. The infinite-volume ground-state energy level, which is shown as red horizontal band, is essentially identical to the finite-volume energies. 

In that way we have determined the infinite-volume ground-state energies for all four different pairs of pion masses and lattice spacings. We present them in Fig.~\ref{fig:scattAnalys} (right) together with a chiral extrapolation to the physical pion mass (lattice discretization errors are ignored at the moment). This leads to an infinite-volume binding energy at the physical pion mass $ m_{\pi,\textrm{phys}}=135\,\textrm{MeV} $ of
\begin{equation}
	\Delta E_0 (m_{\pi,\textrm{phys}}) = (-103 \pm 8 )\,\textrm{MeV}.
\end{equation}
This result is slightly smaller, but still consistent with our previous result $ (-128 \pm 24 \pm 10)\,\textrm{MeV} $ from Ref.~\cite{Leskovec:2019ioa}, where we have used scattering operators only at the sink. Note that further technical details of our ongoing study were presented at the same conference \cite{Wagner:2022bff}.


\section*{Acknowledgments}
We acknowledge useful discussions with Luka Leskovec.\\
M.W. and M.P. acknowledge support by the Deutsche Forschungsgemeinschaft (DFG, German Research Foundation) $ - $ project number 457742095. M.W. acknowledges support by the Heisenberg Programme of the Deutsche Forschungsgemeinschaft (DFG, German Research Foundation) $ - $ project number 399217702. J.F. is financially supported by the H2020 project PRACE 6-IP (GA No.~82376) and by the EuroCC project (GA No.~951732) funded by the Deputy Ministry
of Research, Innovation and Digital Policy and the Cyprus Research and Innovation Foundation and the European High-Performance Computing Joint Undertaking (JU) under grant agreement No.~951732. S.M. is supported by the U.S. Department of Energy, Office of Science, Office of
High Energy Physics under Award Number DE-SC0009913. We thank the RBC and UKQCD collaborations for providing their gauge field ensembles \cite{Aoki:2010dy,Blum:2014tka}. We thank the MILC collaboration for sharing their gauge link ensembles \cite{MILC:2012znn}. Part of the results were obtained using Cyclone High Performance Computer at The Cyprus Institute, under the preparatory access with id \texttt{p054}. Calculations were conducted on the GOETHE-HLR and on the FUCHS-CSC high-performance computers of the Frankfurt University. We would like to thank HPC-Hessen, funded by the State Ministry of Higher Education, Research and the Arts, for programming advice.
This research also used resources of the National Energy Research Scientific Computing Center (NERSC), a U.S.~Department of Energy Office of Science User Facility operated under Contract No.~DE-AC02-05CH11231, as well as resources at the Texas Advanced Computing Center that were part of the Extreme Science and Engineering Discovery Environment (XSEDE), which was supported by National Science Foundation grant number ACI-1548562.

%

\end{document}